\documentclass[aps,twocolumn,showpacs,superscriptaddress,amsmath,amssymb,amsfonts,floatfix,longbibliography]{revtex4-1}

\usepackage[T1]{fontenc} 
\usepackage{graphicx}
\usepackage{float}
\usepackage{dcolumn}
\usepackage{bm}
\usepackage{microtype}
\usepackage{xfrac}
\usepackage{gensymb}
\usepackage[most]{tcolorbox}
\usepackage{xcolor}
\usepackage{multirow}
\usepackage{enumitem}
\usepackage{natbib,hyperref}
\usepackage{graphicx}
\usepackage{ulem}

\hypersetup{
	colorlinks=true, %set true if you want colored links
	citecolor=blue,  %choose some color if you want links to stand out
}

\begin{document}

\title{Impact of synthesis method on the structure and function of high entropy oxides}

\author{Mario U. Gonz\'{a}lez-Rivas}

\affiliation{Department of Physics \& Astronomy, University of British Columbia, Vancouver, BC V6T 1Z1, Canada}
\affiliation{Stewart Blusson Quantum Matter Institute, University of British Columbia, Vancouver, BC V6T 1Z4, Canada}

\author{Solveig S. Aamlid} 

\affiliation{Stewart Blusson Quantum Matter Institute, University of British Columbia, Vancouver, BC V6T 1Z4, Canada}

\author{Megan R. Rutherford} 
\affiliation{Department of Physics \& Astronomy, University of British Columbia, Vancouver, BC V6T 1Z1, Canada}
\affiliation{Stewart Blusson Quantum Matter Institute, University of British Columbia, Vancouver, BC V6T 1Z4, Canada}

\author{Jessica Freese} 
\affiliation{Department of Physics and Engineering Physics, University of Saskatchewan, Saskatoon, SK S7N 5E2, Canada}

\author{Ronny Sutarto} 
\affiliation{Canadian Light Source, Saskatoon, Saskatchewan S7N 2V3, Canada}

\author{Ning Chen}
\affiliation{Canadian Light Source, Saskatoon, Saskatchewan S7N 2V3, Canada}

\author{Edgar E. Villalobos-Portillo}
\affiliation{European Synchrotron Radiation Facility, 38043 Grenoble Cedex 9, France}

\author{Hiram Castillo-Michel}
\affiliation{European Synchrotron Radiation Facility, 38043 Grenoble Cedex 9, France}

\author{Minu Kim}
\affiliation{Max Planck Institute for Solid State Research, Heisenbergstrasse 1, 70569 Stuttgart, Germany}

\author{Hidenori Takagi}
\affiliation{Max Planck Institute for Solid State Research, Heisenbergstrasse 1, 70569 Stuttgart, Germany}
\affiliation{Department of Physics, University of Tokyo, Bunkyo-ku, Hongo 7-3-1, Tokyo 113-0033, Japan}
\affiliation{Institute for Functional Matter and Quantum Technologies, University of Stuttgart, 70550 Stuttgart, Germany}

\author{Robert J. Green} 
\affiliation{Department of Physics and Engineering Physics, University of Saskatchewan, Saskatoon, SK S7N 5E2, Canada}
\affiliation{Stewart Blusson Quantum Matter Institute, University of British Columbia, Vancouver, BC V6T 1Z4, Canada}

\author{Alannah M. Hallas}
\email[Email: ]{alannah.hallas@ubc.ca}
\affiliation{Department of Physics \& Astronomy, University of British Columbia, Vancouver, BC V6T 1Z1, Canada}
\affiliation{Stewart Blusson Quantum Matter Institute, University of British Columbia, Vancouver, BC V6T 1Z4, Canada}

\begin{abstract}
\smallskip
\begin{center}
{\normalsize\textbf{ABSTRACT}}\\    
\end{center}

The term sample dependence describes the troublesome tendency of nominally equivalent samples to exhibit different physical properties. High entropy oxides (HEOs) are a class of materials where sample dependence has the potential to be particularly profound due to their inherent chemical complexity. In this work, we prepare a spinel HEO of identical nominal composition by five distinct methods, spanning a range of thermodynamic and kinetic conditions: solid state, high pressure, hydrothermal, molten salt, and combustion syntheses. By structurally characterizing these five samples across all length scales with a variety of x-ray methods, we find that while the average structure is unaltered, the samples vary significantly in their local structures and their microstructures. The most profound differences are observed at intermediate length scales, both in terms of crystallite morphology and cation homogeneity. As revealed by x-ray fluorescence microscopy ideal cation homogeneity is achieved only in the case of combustion synthesis. These structural differences in turn significantly alter the observed functional properties, which we demonstrate via characterization of their magnetic response. While ferrimagnetic order is retained across all five samples, the sharpness of the transition, the size of the saturated moment, and the coercivity all show marked variations with synthesis method. We conclude that the chemical flexibility inherent to HEOs is complemented by strong synthesis method dependence, providing another axis along which to optimize these materials for a wide range of applications.

\end{abstract}

\maketitle

\section*{Introduction}

Sample-to-sample variability of physical properties is a well-known reality in materials science. Such sample dependence in nominally equivalent specimens can have a multitude of origins including differences in the size or morphology of grains, the density of point defects, and the presence of extended defects. Accordingly, sample dependence is most profound when comparing specimens produced by different synthetic methods, which, due to various thermodynamic and kinetic factors, can produce samples that differ dramatically at the microstructural and even atomic level~\cite{Saidaminov-processing}. These subtle structural differences frequently lead to profound modifications in functional properties. Consequently, the sources of sample dependence must be understood and controlled before a material can be deployed in real world applications.

\begin{figure*}[htb]
  \centering
  \includegraphics[width=17cm]{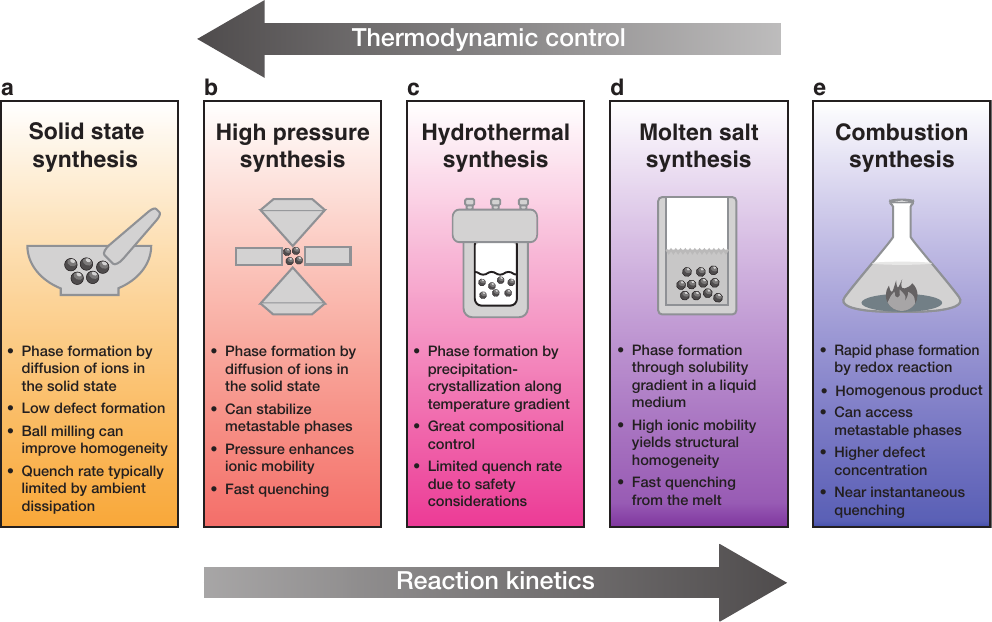}
  \caption{In this study, the HEO spinel (Cr,Mn,Fe,Co,Ni)$_3$O$_4$ has been synthesized by five methods: \textbf{(a)} solid state synthesis, \textbf{(b)} high pressure synthesis, \textbf{(c)} hydrothermal synthesis, \textbf{(d)} molten salt synthesis, and \textbf{(e)} combustion synthesis. Each of these synthetic methods gives rise to varying degrees of thermodynamic and kinetic conditions, and the methods are approximately sorted according to increasing kinetics and decreasing thermodynamic control, from left to right. Our characterization reveals that the most prominent differences across these five nominally equivalent samples occur at the local and microstructural levels, which, in turn, also modify their magnetic properties.}
  \label{Methods}
\end{figure*}

The compositional complexity of HEOs, which are crystalline materials where multiple metal cations share a lattice site~\cite{rost2015entropy,aamlid2023understanding,Brahlek-Whatsinaname-review,musico2020emergent,Sarkar-HEOs-fundamentals}, provides manifold opportunities for sample-to-sample variation. First, HEOs are susceptible to the usual origins of sample dependence in oxides, including grain size, oxygen nonstoichiometry, and other defects~\cite{gu2022engineering,yuan2023role,wang2023high,zhang2023defect}. Beyond these, an HEO specific source of sample dependence is the extent to which the cation distribution meets the perfectly random limit. Depending on synthesis conditions, clustering and short range order~\cite{Page-SRO-PyrochloreHEO,Page-SRO-spinels} can cause a significant departure from the ideal configurational entropy~\cite{chellali2019homogeneity,buckingham2023synthetic}. The forced sharing of a lattice site also makes HEOs prone to local distortions of their oxygen environment~\cite{RostEXAFS,Zhang-long-range-AFM-HEO}, and these too could differ between samples. Exacerbating the potential for sample dependence in HEOs is the sheer number of methods that are employed in their synthesis~\cite{musico2020emergent}. Understanding how synthesis method influences the properties of HEOs is an essential step in their eventual use for applications that require fine-tuning of functional properties.

In this work, we investigate the impact of sample preparation method on the structure and function of HEOs. We prepare the spinel HEO (Cr,Mn,Fe,Co,Ni)$_3$O$_4$ by five distinct synthesis methods, each of which offers different levels of thermodynamic and kinetic control. We structurally characterize these five nominally identical samples across all length scales, finding that while the average spinel structure is maintained, there are significant differences in chemical homogeneity, microstructure, and site selectivity. These structural differences in turn yield samples that vary significantly in the manifestation of their magnetic order, including the sharpness of the transition and the magnitude of the saturated moment. Sample preparation method therefore emerges as yet another control knob for the optimization of HEO functional properties.

\section*{Results and Discussion}
\subsection{Synthesis}

Our investigation of sample dependence in HEOs features the spinel (Cr,Mn,Fe,Co,Ni)$_3$O$_4$, which is amenable to a broad range of synthesis methods. Previous works on this material have established that it is ferrimagnetic~\cite{mao2019facile,musico2019tunable,sarkar2022comprehensive,johnstone2022entropy} with strong cation site selectivity between its tetrahedral and octahedral sublattices~\cite{sarkar2022comprehensive,johnstone2022entropy}, and possesses properties of interest for electrochemical and catalytic applications \cite{Talluri-supercapacitors, Huang-lithiation-mechanism, Nguyen-Delithiation, Wang-anode-spinel, Sarkar-HEOs-fundamentals}. We synthesized polycrystalline samples of the spinel HEO, all with identical nominal composition, by five distinct methods: solid state, high pressure, hydrothermal, molten salt, and combustion syntheses. These five methods span a range of thermodynamic and kinetic conditions, summarized in Fig.~\ref{Methods}, along with key characteristics that might alter the properties of an HEO. Routes with fast reaction kinetics are expected to yield homogeneous cation distributions but present more defects. Conversely, slow reaction kinetics may lead to cation clustering~\cite{aamlid2023understanding} but lower defect concentrations. The phase formation mechanism, reaction temperature, and quench rate will all influence the structural characteristics of the resulting crystalline material. We therefore characterize the structures of our five samples across all length scales.

\begin{figure*}[tb]
  \centering
  \includegraphics[width=17cm]{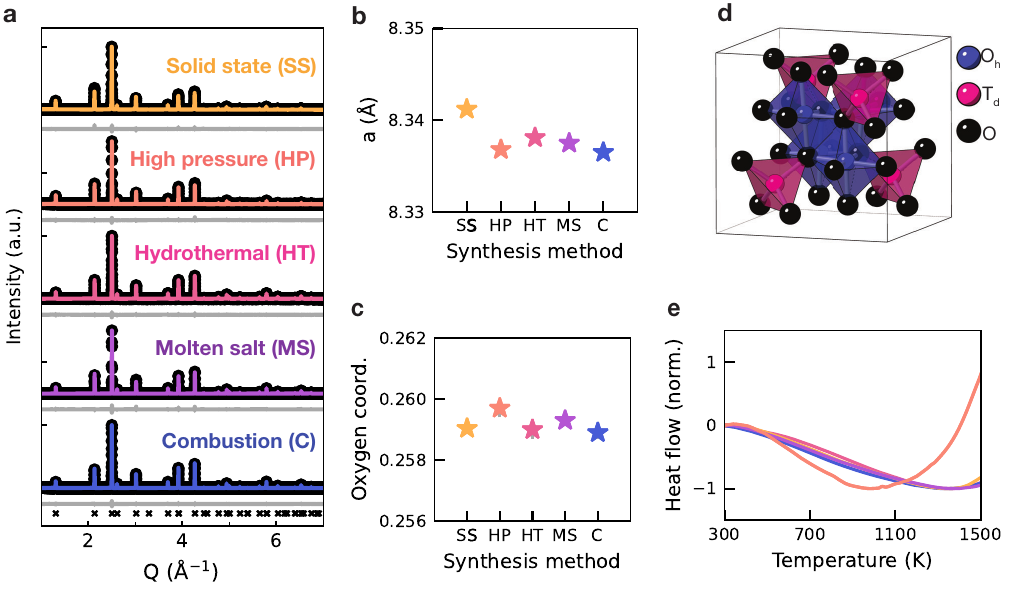}
  \caption{Synthesis method dependence of the average structure. \textbf{(a)} Rietveld refinements for each of the (Cr,Mn,Fe,Co,Ni)$_3$O$_4$ samples prepared by the five different synthesis methods, as labeled. All samples present a cubic spinel phase, with the allowed Bragg peak positions marked by the black crosses. Variation in the \textbf{(b)} refined lattice parameter and \textbf{(c)} refined oxygen coordinate according to synthesis method. The latter corresponds to a modulation of the volume of the local coordination environment between tetrahedral and octahedral lattice sites. Both parameters are highly stable across all synthesis methods. Error bars on these refined values are in the fourth decimal place and are smaller than the symbols. \textbf{(d)} The spinel structure is composed of tetrahedral (light pink) and octahedral (dark blue) lattice sites with the oxygen positions shown in black. Atoms are omitted from the unit cell to highlight the key structural motifs. \textbf{(e)} Normalized differential scanning calorimetry (DSC) curves for spinel HEO samples obtained through various methods. The curve for high pressure synthesis is an outlier, suggestive of metastable configurations being captured through this route.}
  \label{Refinements}
\end{figure*}

\subsection{Average structure}

Powder x-ray diffraction (XRD) patterns for each of our five HEO spinel samples are shown in Fig. \ref{Refinements}(a) and shown in an expanded view in Fig.~S2 of the Supporting Information. All five methods produce samples of high crystalline quality. In the cases of combustion, molten salt, high pressure, and hydrothermal syntheses, no impurity phases were detected, while the solid state sample contains a minor (2\%)  rock salt impurity, as reported in Ref.~\cite{johnstone2022entropy}. The refined cubic lattice parameter (space group $Fd\overline{3}m$) and oxygen coordinate for each sample are plotted as a function of synthesis method in Fig.~\ref{Refinements}(b,c). Both of these quantities show remarkable stability across the five samples: the lattice parameter varies by less than 0.005~\AA\ while the oxygen coordinate differs by less than 0.002 across the series. This parameter relates the volume of the octahedral and tetrahedral coordination environments, with smaller values giving rise to enlarged octahedra and contracted tetrahedra. The stability of the coordination environment, despite the local structure differences that will be highlighted below, suggests that steric effects in the spinel HEO are governed by the law of averages.

\begin{figure*}
  \centering
  \includegraphics[width=14 cm]{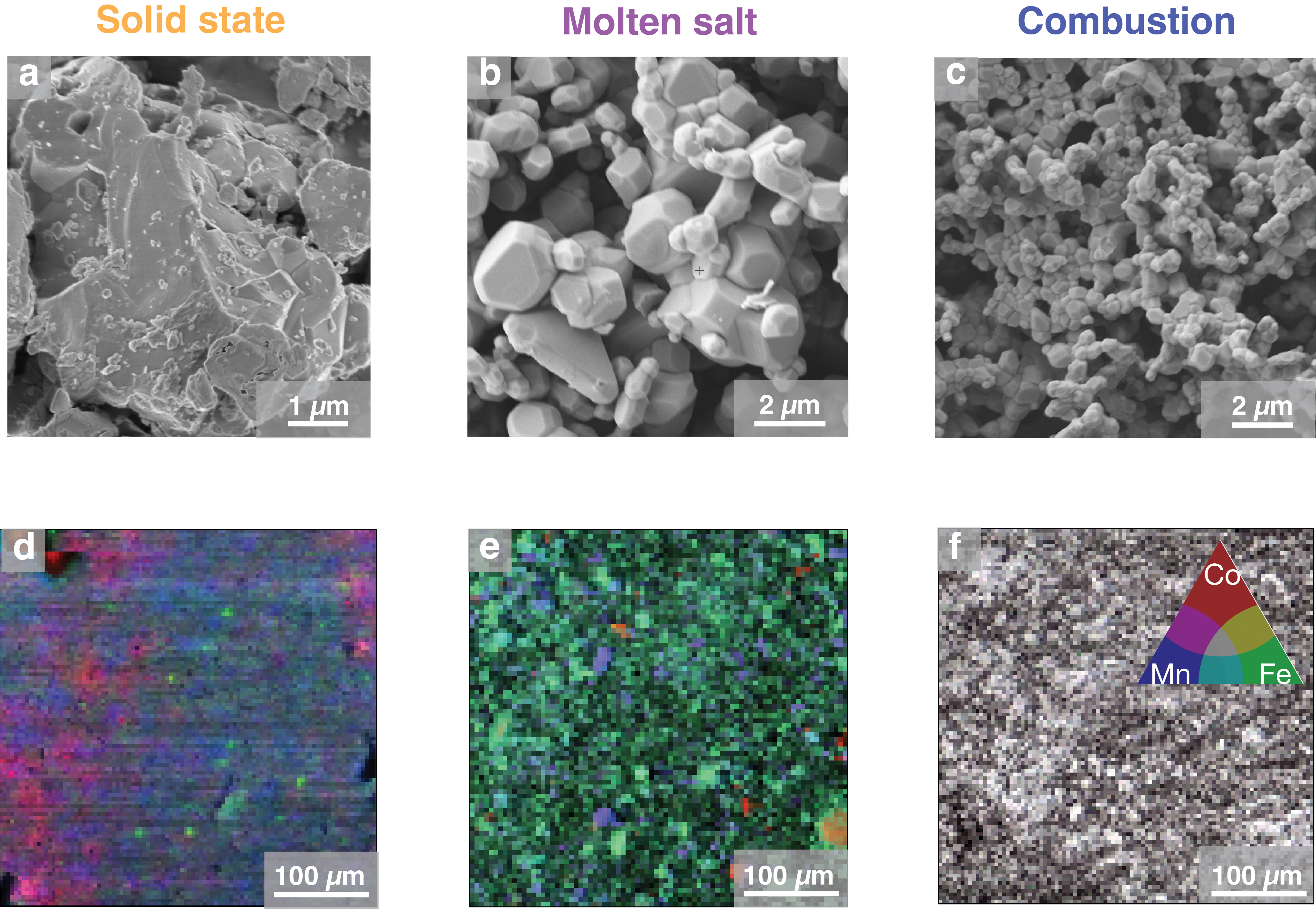}
  \caption{Synthesis method dependence of the spinel HEO's microstructure. \textbf{(a-c)} Scanning electron micrographs of spinel HEO samples obtained from three distinct synthesis methods showing striking differences in microstructure. The solid state sample in panel \textbf{(a)} shows highly irregular particle shapes, with sizes starting at 5 $\mu$m, while the  molten salt sample in \textbf{(b)} is primarily composed of micron-sized crystals with a rounded octahedron crystal habit and the combustion sample in \textbf{(c)} forms visibly porous sub-$\mu$m structures composed of round particles. \textbf{(d-f)} $\mu$-XRF RGB maps showing varying compositional homogeneity for spinel HEO samples prepared via different methods. Deviations from ``neutral'' colors (gray, white) indicate local nonstoichiometry and an excess of one or more elements, as defined by the colors in the triangular legend. Black spots in panels \textbf{(e,f)} correspond to regions devoid of material, as the molten salt and combustion samples were run as powders, while the solid state sample in \textbf{(d)} was run as a pellet. The solid state sample \textbf{(d)} has a significant degree of chemical inhomogeneity, with Co and Mn-rich regions tens of $\mu$m wide, along with smaller, roughly 1~$\mu$m Fe-rich regions. The green tinge throughout the majority of the map for the molten salt sample in \textbf{(c)} indicates the presence of excess Fe. Mn-rich (blue) and Co-rich (red) regions can also be observed, along with a slightly Mn-deficient (orange) spot on the lower right of the map. The scale of these regions is in the tens of $\mu$m. The white and gray tones throughout the map for the combustion sample in \textbf{(d)} indicate that this sample is exceptionally homogeneous within the resolution of the instrument.} 
  \label{XRF}
\end{figure*}

We can also use the powder XRD data to assess the crystallinity of the different samples. Supporting Figure S1 provides a closer look at the (311) Bragg reflection as a function of synthesis method. The solid state sample stands out due to having the largest full width at half maximum (FWHM), indicative of either smaller crystallite sizes or increased inhomogeneity. The former explanation can be excluded here, as among these five methods, solid state should produce the largest structural grains. Instead, we ascribe the broader peaks as originating from a larger distribution of lattice parameters due to chemical inhomogeneity such as cation segregation. The other four samples exhibit exemplary crystallinity with minimal broadening and symmetric Bragg peaks. We can therefore conclude that at the \textit{average} structure level, our samples, with the exception of solid state, are largely indistinguishable.

Another key aspect of HEOs is their thermodynamic behavior, making differential scanning calorimetry (DSC) a particularly useful technique to characterize these materials. Fig.~\ref{Refinements}(e) presents DSC measurements for all samples, normalized to their minima. Four of the five samples present nearly indistinguishable behavior, with negative heat flow across the full measured temperature range, consistent with the endothermic behavior expected for an entropy stabilized material~\cite{aamlid2023understanding}. The DSC for each of these samples exhibits a broad minimum centered at approximately 1400~K, marking the temperature scale on which the spinel phase fully decomposes. Only the high pressure sample's response is dramatically different, with its minima reproducibly occurring closer to 1000~K in both pellet and powder form, as shown in Fig. S3(b), ruling out surface effects as a possible origin. This difference in thermal behavior suggests that the application of extreme pressure and the corresponding modified thermodynamic conditions traps the HEO spinel with some metastable cation configurations that cannot be detected at the average structure level. After returning to ambient pressure, these unfavorable cation arrangements remain kinetically trapped until sufficient thermal energy allows them to re-configure. Considering the other four samples, we find that the magnitude of the DSC response varies dramatically, as can be seen in the unnormalized data presented in Fig. S3(a). We ascribe these quantitative differences to microstructural differences, to be discussed next, which allow heat to be transferred more or less effectively across grain boundaries.

\subsection{Microstructure}

Moving one length scale down, we next consider the microstructure of our spinel HEO samples. Microscale differences in particle size and shape in polycrystalline samples can have dramatic effects on a material's properties, making their characterization paramount to a thorough understanding of synthesis dependence. Scanning electron microscopy (SEM) was used to characterize the particle morphology attained with each synthesis method. Representative micrographs for the solid state, molten salt, and combustion syntheses are displayed in Fig.~\ref{XRF}(a-c), as these three samples span the range of phase formation mechanisms. Representative micrographs for the other methods are provided in Fig. S3. These micrographs reveal that our samples have dramatically different morphologies. Solid state and high pressure syntheses result in highly irregular particle shapes, with sizes on the order of 5~$\mu$m. Meanwhile, molten salt and hydrothermal syntheses produce highly faceted, micrometric crystals around 2-5~$\mu$m across. Finally, combustion synthesis results in a highly porous networked structure of nanometric (100-500~nm) particles. As a result of their different grain morphologies, our five spinel HEO samples vary dramatically in their surface areas, which is significant in the context of their catalytic activity.

Beyond morphological differences, microscale variations in chemical homogeneity can also dramatically impact the functional properties of solids. X-ray fluorescence microscopy ($\mu$-XRF) employing synchrotron radiation exploits the quantization of elemental fluorescence lines and couples them with a microscopic spot size to expose variations in the chemical composition of solids at the microscale. $\mu$-XRF is a well-established tool in the fields of toxicology, environmental sciences, and art conservation~\cite{mu-XRF-VanGogh-conservation,mu-XRF-Nanowire-toxicity,mu-XRF-environmental-mine}. Here, we harness its strengths to expose chemical inhomogeneity and nonstoichiometry variations in nominally equivalent spinel HEO samples.

Representative $\mu$-XRF maps for the solid state, molten salt, and combustion samples are shown in Fig.~\ref{XRF}(d-f) on a 400 $\times$ 400 $\mu$m scale. Additional probe regions for each of these three samples, consistent with the findings discussed here, are presented in Fig.~S6. In these maps, colored pixels expose nonstoichiometric regions. Dramatic inhomogeneity is observed for the solid state sample, with spatially extended red and blue regions indicating Co and Mn-rich regions, respectively, over length scales of 10 microns or more. Bright green pixels indicating localized Fe-rich regions are also apparent. These inhomogeneities are consistent with the limitations of the diffusive process for solid state but do still respect the overall expected stoichiometry for the spinel HEO.

Significant chemical inhomogeneities are also apparent in the molten salt sample, which is mainly Fe-rich (green) with localized Mn-rich (blue), Co-rich (red), and Mn-deficient (orange) regions as can be seen in Fig.~\ref{XRF}(e). Molten salt synthesis relies on the constituent cations being similarly soluble so that precipitation from the melt can occur simultaneously. However, the solubilities of $3d$ metal oxides in halide fluxes depend strongly on their ionic radii~\cite{Cherginets-oxidesolubilities-halide-fluxes}, precluding co-precipitation and causing the observed nonstoichiometry and inhomogeneity. 

In stark contrast, the combustion sample exhibits exceptional homogeneity and stoichiometry, with the 400 $\times$ 400 $\mu$m map showing minimal deviations from neutral tones. Therefore, the combustion sample better fulfills the most stringent definitions of HEOs~\cite{aamlid2023understanding}. Its exceptional quality can be traced to the combustion mechanism, which requires an atomically mixed, homogeneous gel, maximizing the randomness of the cation distribution. Our measurements therefore highlight how HEO samples that are indistinguishable in their average structures can differ dramatically at the microscale.

\subsection{Local Structure}

\begin{figure*}
  \centering
  \includegraphics[width=17cm]{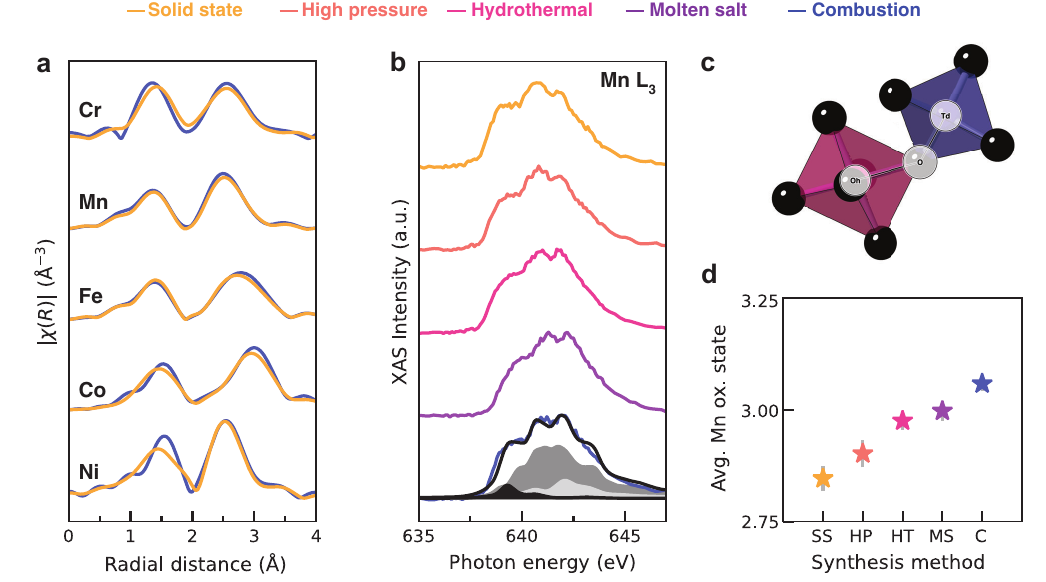}
  \caption{Synthesis method dependence of local structure probed with resonant x-ray techniques. \textbf{(a)} Extended x-ray absorption fine structure (EXAFS) $|\chi(r)|$ data for the solid state and combustion samples measured on the K-edge for each cation. \textbf{(b)} Mn L\textsubscript{3} x-ray absorption spectra (XAS) for all five samples. A representative multiplet calculation is shown for the combustion sample, where black corresponds to tetrahedral Mn$^{2+}$, dark gray to Jahn-Teller distorted octahedral Mn$^{3+}$, and light gray to octahedral Mn$^{4+}$. \textbf{(c)} Fundamental octahedral-tetrahedral corner-sharing unit for the spinel structure. \textbf{(d)} Average Mn oxidation state as a function of synthesis method, obtained from multiplet calculation fits to the entire Mn L\textsubscript{2,3} spectra.}
  \label{XAS-EXAFS}
\end{figure*}

The last length scale to consider is the local regime, corresponding to the first and second-neighbor coordination environments of the various cations, which govern the local crystal fields and the superexchange interactions between tetrahedra and octahedra in the spinel structure. Characterizing the different samples in this regime is crucial as the magnetic properties of transition metal spinel oxides are remarkably sensitive to the distribution of cations across their lattice sites~\cite{DronovaControllingInversion2022}. To this end, we rely on two complementary flavors of core-level spectroscopy: extended x-ray absorption fine structure (EXAFS), and soft x-ray absorption spectroscopy (XAS). 

We collected EXAFS data at the transition metal K-edges for spinel HEO samples synthesized via solid state and combustion syntheses, as these samples represent opposite ends of the thermodynamic spectrum presented in Fig.~\ref{Methods}. The resulting real-space $|\chi(R)|$ data are shown in Fig.~\ref{XAS-EXAFS}(a), with more comprehensive views of the spectra provided in Fig. S5. For each cation, the EXAFS data show a first peak around 1.5~\AA, corresponding to cation-oxygen scattering in the first coordination shell, and a second peak around 2.5~\AA, corresponding to scattering between cations on adjacent tetrahedral and octahedral sites, as shown schematically in Fig.~\ref{XAS-EXAFS}(c). Overall, there is a general trend toward a higher degree of resolved fine structure for the combustion sample, consistent with its higher homogeneity and sample quality as shown in the bulk and microstructure sections. 

Taking each cation in turn, two of the edges (Mn and Fe) are largely identical, pointing to a high degree of similarity in their local chemical environments. For Co, the peak at the first coordination shell shows a clear splitting for the combustion sample that is smoothed out in the case of the solid state sample. This smearing could be ascribed to small variations in the tetrahedral crystal field environment, as consistent with the larger uncertainty in lattice parameter implied by the broader diffraction peaks for the solid state sample. Large differences are found for Ni, where an overall broadening and spectral weight redistribution is observed in the first coordination shell for the solid state sample, compared to combustion. These discrepancies can be largely understood in terms of the solid state sample's NiO impurity, which will result in variations for the first coordination shell while retaining similar distances for the second nearest neighbors. As for Cr, the Cr-O distance appears to be larger in the solid state sample, contrary to the overall trend. This observation suggests that Cr is exerting negative chemical pressure to keep the average size of the octahedral environments constant.

To complement the qualitative local structure picture attained from EXAFS, we now explore our XAS L\textsubscript{2,3} data. Spectra for Cr, Fe, and Ni are largely indistinguishable across all five samples, as shown in Fig. S6. Both Cr$^{3+}$ and Ni$^{2+}$ exclusively occupy octahedral sites due to strong crystal field effects. At the opposite end of the spectrum, Fe$^{3+}$, due to its half-filled $d$-shell, is spherically symmetric and shows no site selectivity, occupying a nearly 50-50 mixture of octahedral and tetrahedral sites in all cases. Likewise, the Co L\textsubscript{2,3} XAS remains fairly unchanged, and is characterized by the presence of Co$^{2+}$ in a tetrahedral configuration.

\begin{figure*}[tb]
  \centering
  \includegraphics[width=17cm]{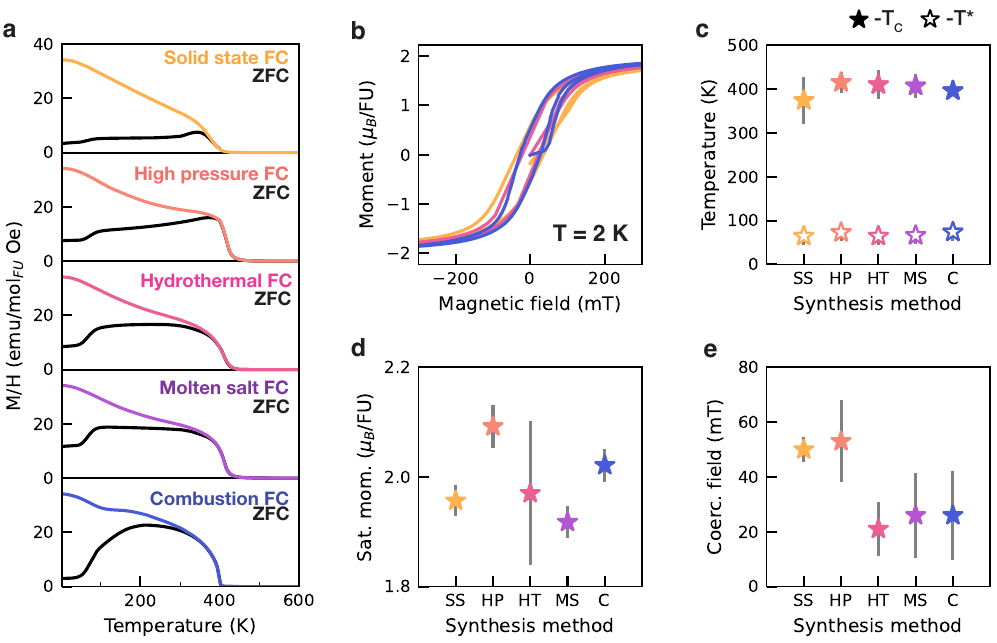}
  \caption{Synthesis method dependence of magnetic properties. \textbf{(a)} Temperature dependence of the magnetic susceptibility, $M$/$H$, measured in an $H=100$~mT field for each of the (Cr,Mn,Fe,Co,Ni)$_3$O$_4$ samples prepared by the five different synthesis methods, as labeled. Each sample undergoes a ferrimagnetic ordering transition close to 400~K, below which there is a bifurcation of the susceptibility measured under zero field-cooled (ZFC) and field-cooled (FC) conditions. \textbf{(b)} Magnetic moment as a function of applied field for selected samples at $T = 2$~K showing soft hysteretic behavior and saturation in fields around $H=200$~mT. The synthesis method dependence of the key magnetic properties, \textbf{(c)} Curie temperature ($T_C$) and depinning temperature ($T*$), \textbf{(d)} saturated moment normalized per formula unit ($\mu_{\text{sat}}$/FU), and \textbf{(e)} coercive field ($H_c$), revealing remarkable consistency across samples despite their micro and local structural differences.}
  \label{Magnetometry}
\end{figure*}

Out of the five cations, Mn presents the richest XAS L\textsubscript{2,3} phenomenology, which although visually subtle, indicates chemically meaningful variations with synthesis method. For their quantification, we used cluster calculation models, shown in Fig. S7. The Mn L\textsubscript{3} spectra presented in Fig.~\ref{XAS-EXAFS}(b) are characterized by three features corresponding to three unique Mn states. The first of these is tetrahedral Mn$^{2+}$ centered at 639 eV, then Jahn-Teller distorted octahedral Mn$^{3+}$ at 641 eV, and finally octahedral Mn$^{4+}$ at 642 eV. The breakdown of these contributions is shown for the combustion sample along with the resulting fit to the data. Variations in the Mn$^{3+}$ and Mn$^{4+}$ relative intensities are observed as a function of reaction kinetics, giving rise to a shift in the average Mn oxidation state, as presented in Fig.~\ref{XAS-EXAFS}(d). As all other cations exhibit stable fixed oxidation states, we can directly calculate the expected Mn oxidation state, assuming perfect spinel stoichiometry, giving 3.3$\overline{3}$+. The fitted values fall short of this expectation, indicating oxygen vacancies that reach the highest concentration in the samples closest to thermodynamic equilibrium. Therefore, although our five samples have nominally equivalent compositions, we can conclude that synthesis method can significantly alter the final cation to oxygen stoichiometry.

Careful consideration of the previously discussed microstructural data can help pinpoint the most likely origin for these variations in cation to oxygen stoichiometry. With the exception of high-pressure, the final step in the synthesis protocol for each method used here is a 24 h annealing in air at 1050 \degree C. One would therefore assume that, in each case, steady state would be achieved in terms of the sample's oxidation/reduction potential and these four samples would have equivalent oxygen stoichiometries. Instead, the observed trend appears to correlate precisely with the microstructure and, in particular, the amount of exposed surface area. For example, combustion synthesis, which yields the most oxidized sample has a highly porous network that promotes a higher degree of oxygen incorporation. In contrast, solid state and high pressure present very limited surface areas, and can only incorporate oxygen at the surface. Hydrothermal and molten salt samples form as micrometric crystals, giving them an intermediate surface area thanks to their exposed crystal facets. Overall, we conclude that the formation of microstructures with larger, exposed surface areas, drives a closer to ideal cation to oxygen stoichiometry in the synthesis of spinel HEO materials. More generally, in cases where mitigating the presence of oxygen vacancies is desirable, maximizing the material's surface area at the point of the annealing cycle provides a straightforward optimization pathway.

\begin{table*}[htbp]
\caption{Characteristic structural, chemical, and functional features of HEO samples prepared via each synthesis method.}
\label{summary_table}
\begin{tabular}{lccccc}
\hline
Synthesis method & Structure         & Stoichiometry                                                                    & Homogeneity      & Magnetism                                                                & Other features                                                                          \\ \hline
Solid state      & Least crystalline & \begin{tabular}[c]{@{}c@{}}Best compositional \\ control\end{tabular}            & Least homogeneous & \begin{tabular}[c]{@{}c@{}}Broadest transition,\\ lowest $T_C$    \end{tabular}                                                  & Most scalable                                                                           \\ \hline
High pressure    & –                 & –                                                                                & –                & \begin{tabular}[c]{@{}c@{}}Largest coercivity,\\ highest $T_C$\end{tabular} & Least scalable                                                                          \\ \hline
Hydrothermal     & –                 & –                                                                                & –                & Smallest coercivity                                                      &  \begin{tabular}[c]{@{}c@{}}Lowest temperature \\ phase formation \end{tabular}                                                                                 \\ \hline
Molten salt      & Largest crystals  & \begin{tabular}[c]{@{}c@{}}Worst compositional \\ control\end{tabular}           & –                & –                                                                        & \begin{tabular}[c]{@{}c@{}}Most faceted crystals, \\ scalable\end{tabular}              \\ \hline
Combustion       & Most crystalline  & \begin{tabular}[c]{@{}c@{}}Closest to ideal oxygen \\ stoichiometry\end{tabular} & Most homogenous  & Sharpest transition                                                      & \begin{tabular}[c]{@{}c@{}}Largest surface area, \\ fastest reaction times\end{tabular} \\ \hline
\end{tabular}
\end{table*}

\subsection{Magnetic Properties}

Having thoroughly characterized the structural differences among our five samples, we now consider whether the microstructural and nanoscale differences we detect have a significant impact on functional properties. Magnetic susceptibility, a probe of the average magnetic response, is an ideal technique to address this question. The magnetic behavior of individual atoms is strongly influenced by their local environments while bulk magnetic properties are often sensitive to the level and type of defects as well as microstructure.

We first turn our attention to the magnetic susceptibility ($M$/$H$) as a function of temperature for each synthesis method, shown in Fig.~\ref{Magnetometry}(a). All five samples order ferrimagnetically with Curie temperatures, $T_C$, close to 400~K, as marked by a sharp increase in the susceptibility and splitting between the field-cooled (FC) and zero-field-cooled (ZFC) measurement conditions. A second feature around 70~K, denoted $T^*$, manifests as an inflection in the ZFC susceptibility and marks the temperature at which there is enough thermal energy to overcome the pinning of magnetic domains~\cite{johnstone2022entropy}. Qualitatively, a continuous evolution in the overall shapes of the susceptibility curves is observed as a function of synthesis method and appears to follow the ranking of the methods in Fig.~\ref{Methods} according to their thermodynamic and kinetic control. In particular, one can notice that going from solid state to combustion synthesis, the transition becomes progressively sharper, the inflection at $T^*$ in both the ZFC and FC curves becomes sharper, and the bifurcation moves to lower temperatures. These differences appear to correlate with the level of chemical homogeneity, as observed in the $\micro$-XRF measurements across the five samples. 

We can next look to the magnetization as a function of applied field, where hysteresis loops are shown in Fig.~\ref{Magnetometry}(b) for select samples. All samples present similar, soft ferrimagnetic behavior, with small coercive fields and saturated moments that are close to 2~$\mu_B$ per formula unit (FU). This relatively small saturated moment originates from the partial sublattice cancellation due to the opposing direction of the moments on the tetrahedral and octahedral sublattices in the ferrimagnetic ordered state. Interestingly, all five samples have an unconventional feature where, starting from the ZFC condition, the moment is slower to grow as compared to once the moment is saturated and swept back and forth, such that the virgin curve lies outside the hysteresis loop in a small range of applied fields. This is likely a feature related to the growth of magnetic domains, which grow as they are polarized by the applied field and are subsequently switched more easily.

To quantify the magnetic differences across our five samples, we can consider several key features. First, their Curie temperatures (defined here as the minimum of the derivative of the susceptibility with respect to temperature) hover around 400~K, as presented in Fig.~\ref{Magnetometry}(c). The error bars are defined by the FWHM of the peak in the derivative, a measure of the sharpness of the transition. Across the series, $T_C$ varies from 374~K (solid state) to 415~K (high pressure) while the domain depinning at $T^*$ varies by less than 10~K across the five samples. Two additional quantifiable metrics come from the magnetization, which are the saturated moment and the coercive field, shown in Fig.~\ref{Magnetometry}(d,e). The saturated moment should be particularly sensitive to the individual elements' oxidation states and site occupancies, which will in turn dictate the moment on each atom. No systematic trend is apparent and the data cannot, for example, be correlated with the varying Mn oxidation state. One intriguing possibility that has not yet been investigated in HEOs is that the different samples have different fractions of non-ordered glassy or dynamic moments. The coercive field is small in all cases and is observed to correlate with grain size where domain reorientation is more easily achieved in the samples with smaller crystallites. 

To contextualize the spinel HEO’s magnetism, it is particularly interesting to consider the large subset of spinels that are composed of these five transition metals: with the exception of antiferromagnetic Co$_3$O$_4$, spinels composed of these transition metals are uniformly ferrimagnetic. However, unlike the five samples presented here, there are dramatic differences in their ordering temperatures (ranging from 43~K in Mn$_3$O$_4$ to a remarkable 868~K in NiFe$_2$O$_4$), their saturated moments, and their coercive fields. This is because the superexchange in each material depends strongly on the particular filling of the $d$ orbitals and therefore depends on the chemical identity of the elements at each site. In the HEO spinel, with five such transition metals distributed across two sites, it is remarkable that the microstructural and local scale differences resolved here do not more significantly modify the ordering temperature. However, the sharpness of the transition, the size of the saturated moment, and the coercivity, all of which are important functional properties, do show significant dependencies on synthesis method.

\subsection{Discussion}

Taken together, our results show that an HEOs structure and function are intimately tied to the synthesis method used to prepare it, with the most profound differences occurring at the microstructural level. For any future study, the choice of synthesis method is therefore a key study design step that must be considered. For instance, studies that seek to understand the fundamental properties of HEOs may prefer methods such as combustion that produce the most homogeneous cation distribution, which is assumed in the ideal definition of an HEO~\cite{Brahlek-Whatsinaname-review,aamlid2023understanding}. In other cases, particularly when considering applications, other features such as scalability, reproducibility, and energy cost may take precedence. These factors must be weighed in relation to other structural and functional optimizations that the different synthesis methods naturally impart. The key features of each method are summarized in Table~\ref{summary_table}.

Reproducibility is another area of concern, particularly in a rapidly progressing field such as HEOs. Four of the five methods described here (high pressure, hydrothermal, molten salt, and combustion) require tightly controlled synthesis conditions, which we expect would minimize sample variability. The poorest reproducibility in our experience occurs with solid state synthesis due to kinetic limitations which are highly sensitive to the degree of pre-homogenization of the oxide precursors. For methods based on mineralizers and fluxes, where relative constituent solubilities have long been established as critical to controlled crystal growth~\cite{BerryToolsTricks2024}, careful reporting of even minute experimental parameters (\textit{e.g.} deionized/distilled water conductivities, detailed heating cycles) is critical to ensure reproducibility of results and open up avenues for further optimization.

Along similar lines, the findings presented here also naturally raise the question of the consistency between high entropy samples prepared within a single growth method while varying synthesis parameters. While we expect that variations within a synthesis method would generally be smaller than those between synthesis methods, there can be no doubt that the former is also susceptible to sample dependence. For instance, in the case of solid state with the slowest reaction kinetics, there is a significant sensitivity to the level of pre-homogenization. In the case of combustion synthesis, negligible differences are expected in cation homogeneity due to the complex formation between cations and fuel~\footnote{Provided proper fuel choice}. Further optimization might be achieved through innovative thermal cycles, applied electromagnetic fields, controlled atmospheres or additives, which could disrupt phase stability, crystal growth anisotropies, and functional properties in HEOs, based on findings for conventional materials~\cite{Wu-MagField-Growth, peng2000shapeCdSE, ChenScienceBlackTiO2}.

Chemical additives, in particular, warrant their own discussion. In this study, we have endeavored to look beyond method-dependent variables (\textit{e.g.} the choice of the fuel, flux, or mineralizer in combustion, molten salt, or hydrothermal syntheses) and instead isolate a method's defining features, such as the high porosity of combustion powders being a direct consequence of gas evolution during the reaction, which will occur for all fuel choices to varying degrees, or the formation of relatively large crystals in molten salt or hydrothermal, where fluxes and mineralizers enable transport. These additives mold the material's properties across all length scales. Going forward, a particularly interesting avenue to explore is the thoughtful inclusion of additional chemicals or solvent replacements, which can help alleviate the limitations of each method (such as chemical inhomogeneity), or provide additional control (for example, by restricting nucleation to provide monodisperse particle sizes).

\section*{Conclusion}

We have shown that the synthesis method employed in preparing an HEO can result in profound microstructural differences, along with important variations in the local structure, and that these effects are masked in the average structure. Specifically, by preparing an HEO spinel with nominally identical chemical composition by five different synthetic methods, we find vastly different grain morphologies, oxygen stoichiometries, local distortions, and - most importantly for an HEO - chemical homogeneities across length scales. These structural differences yield important modifications to functional properties, which we demonstrate here taking magnetic properties as an example. The effects of synthesis method on HEOs will likely be even more profound when considered in the context of other functional properties, in particular electrochemical activity and lithium storage, which conventionally depend strongly on local and microstructural defects. 

Whilst our results were obtained for an HEO with the spinel structure, the underlying effects are kinetically or chemically driven, and hence broadly relevant to all families of high entropy ceramics. In many cases, our findings are linked to the fundamental phase formation mechanisms, such as the chemical inhomogeneity observed for methods that rely on coprecipitation (hydrothermal and molten salt), compared to those that rely on complex formation (combustion). With these considerations, our findings provide insights into the likely behavior of HEOs prepared using synthesis methods not discussed in this article, but that rely on similar chemistry to the ones presented here. This work establishes synthesis method as an additional layer in the design of HEOs, beyond compositional variations.

\section*{Experimental}

\subsection{Synthesis of spinel HEO powders}

Polycrystalline spinel HEO samples with nominal composition (Cr,Mn,Fe,Co,Ni)$_3$O$_4$ were obtained by several synthesis routes, namely solid state, high pressure, hydrothermal, molten salt, and combustion syntheses. The different protocols are described in detail below. In all cases, the samples were quenched from their final annealing temperatures by direct extraction from the furnace.\\

\noindent \textbf{Solid state synthesis}: The solid state sample was prepared following the same method used in Ref.~\cite{johnstone2022entropy}. Stoichiometric amounts of the binary oxides — Cr$_2$O$_3$, MnO$_2$, Fe$_2$O$_3$, Co$_3$O$_4$, and NiO — were initially mixed and manually ground in a mortar and pestle until visibly homogeneous. The mixture was then mechanically homogenized in a ball mill using zirconia milling balls and ethanol for three hours. The resulting mixture was pelleted and slowly heated to 1050 $^\circ$C over 11 hours, with a dwelling time of 36 hours.\\

\noindent \textbf{High pressure synthesis}: For the high pressure synthesis, the unannealed combustion product (procedure described below) was used as a precursor. The powder was mixed with 4 wt. \% KClO$_4$, which acts as an oxidizer due to the slight oxygen deficiency of the unannealed combustion product. The mixture was manually compressed in a Pt capsule and pressurized to 10 GPa in a Walker-type multi-anvil module. The sample was heated to 1000 \degree C for 1 hour before thermal quenching down to room temperature and slow decompression. The resulting dense pellet was taken out of the capsule and the remaining KCl salt from the oxidizer washed away before further analysis. Further details are provided in Ref.~\cite{Aamlid-APL-HP-HEOs}, which explores the effect oxidizer and pressure conditions have on the phase stability of the spinel HEO studied here.\\

\noindent \textbf{Hydrothermal synthesis}: The hydrothermal synthesis method developed here exploits the relatively low decomposition temperatures of metal hydroxides. In a typical procedure, the corresponding metal nitrates were dissolved in a minimal amount of deionized water with a conductivity of 18.2 M$\Omega\cdot$cm. Separately, KOH corresponding to a 5 M concentration in a 12 mL volume was placed in a 23 mL vial. The metal nitrate solution was then suddenly added to the potassium hydroxide, resulting in a vigorous coprecipitation reaction. The precipitates were then dispersed using a sonicator for 10 minutes, and the volume was adjusted to 13 mL by adding an appropriate amount of deionized water. The vial was then sealed inside a Parr Instruments 4749 acid digestion vessel, and heat treated at 180 \degree C for 36 h. The resulting product was washed with deionized water five times, and then vacuum filtered.

At this step, we obtained a dark brown powder, which x-ray diffraction revealed to be a phase mixture. The diffraction pattern was dominated by heavily oriented string-like FeK$_3$O$_4$ and H$_2$NiO$_2$ crystals, coexisting with a nanocrystalline compositionally complex spinel. Upon grinding and annealing this mixture at 1050 \degree C for 24 h, it crystallized into a phase-pure spinel HEO powder. No traces of potassium were detected in the final product, with the most likely outcome being its evaporation at the annealing temperatures.\\

\noindent \textbf{Molten salt synthesis}: For the molten salt synthesis procedure, binary oxides (Cr$_2$O$_3$, MnO$_2$, Fe$_2$O$_3$, CoO, and NiO) and KCl were used as starting reagents. Stoichiometric amounts of the oxide precursors were first weighed and mixed until homogenized. Then, KCl was added in a 1:1 (Cr,Mn,Fe,Co,Ni)$_3$O$_4$
molar ratio and the resulting mixture was ground until a homogeneous orange-brown fine powder was obtained. The powder was then transferred to an alumina crucible and fired at 1050 \degree C for 5 h, followed by a slow descent to 850 \degree C over 12 h, at which point the product was air-quenched by direct extraction from the furnace. The resulting powder was then washed ten times with deionized water to dissolve the KCl flux and annealed at 1050 \degree C for 24 h to eliminate remnant HEO rocksalt impurities. At this stage, a phase-pure spinel HEO sample was obtained.\\

\noindent \textbf{Combustion synthesis}: The combustion synthesis method was adapted from that described in Ref.~\cite{spinel-HEO-discovery}. Here, metal nitrates — Cr(NO$_3$)$_3\cdot$9H$_2$O, Mn(NO$_3$)$_2\cdot$4H$_2$O, Fe(NO$_3$)$_3\cdot$9H$_2$O, Co(NO$_3$)$_2\cdot$6H$_2$O, and Ni(NO$_3$)$_2\cdot$6H$_2$O— were used as oxidants for the combustion process, with glycine acting as the fuel and reducing agent. The reaction follows the general structure:

\begin{math}
    Me^{\nu +}(\text{NO}_3)_{\nu} + \frac{5}{9}\phi \nu \text{C}_2\text{H}_5\text{NO}_2 +\frac{5}{4} \nu (\phi-1) \rightarrow Me^{\nu +}\text{O}_{\nu/2} + \left(\frac{5\phi+9}{18} \right) \nu \text{N}_2 + \frac{25}{18}\nu \phi \text{H}_2 + \frac{10}{9} \nu \phi \text{CO}_2
\end{math}

For the material studied in this paper, an oxidants to reducers ratio $\phi=1$ was used, and the following protocol was followed to obtain our samples. First, a stoichiometric mixture of the metal nitrates was dissolved in a minimal amount of deionized water. After stirring the resulting solution for 10 minutes, the appropriate amount of glycine was added to the mixture, which was then heated to 80 \degree C. Upon reaching this temperature, the solution changed color from an orange-red to a dark brownish-red hue, indicating complexation of the metal cations with the glycine molecules. This mixture was then condensed at 80 \degree C for 24 h to obtain a dark red, viscous xerogel. 

The xerogel was then placed directly in a furnace preheated to 650 \degree C and left to react for an hour. A fluffy and extremely porous needle-like black powder was obtained, comprised of spinel HEO nanoparticles. This powder was then ground and annealed at 1050 \degree C for 5 h yielding a fine black powder. Conducting the reaction on top of a hot plate instead of a furnace produced identical results. Likewise, annealing for 24 h had no impact on the material's crystal structure, crystallinity, or magnetic properties.

\subsection{Powder x-ray diffraction}

Powder x-ray diffraction measurements were carried out at room temperature using a Bruker D8 Advance diffractometer equipped with a copper source and Johansson monochromator, resulting in a monochromatic beam wavelength of $\lambda_{K_{\alpha_1}}$ = 1.5046 \AA. The instrument is equipped with a LYNXEYE XE-T silicon strip detector that enables it to efficiently filter background fluorescence, which is significant for the elements under study. Measurements were carried out over a range of 1 < $Q$ < 6 \AA$^{-1}$, with a step size of 99.7 n\AA$^{-1}$ only for the molten salt sample, and 78.3 n\AA$^{-1}$ for samples obtained through the other synthesis methods. The smaller step size was required to properly resolve the sharper peaks observed in these samples. Rietveld refinements for all samples were performed using the FullProf Suite \cite{rodriguez1993recent}. In all cases, 5\% of $K_{\alpha_2}$ bleed-through radiation, with a wavelength $\lambda_{K_{\alpha_2}}$ = 1.5444 \AA\ was included in the refinement, accounting for a slight wavevector-dependent peak splitting, which was most evident at larger angles. All measurements were conducted on finely ground powders (sifted through a 400 mesh) so that any variations in the diffraction patterns should be intrinsic to the samples themselves.

\subsection{Differential scanning calorimetry}

Differential scanning calorimetry was carried out between room temperature and 1500~K, using a Netzsch Simultaneous Thermal Analysis (STA) 449 F5 Jupiter instrument. Measurements were conducted under flowing nitrogen with a ramping rate of 20~K/min. All DSC measurements were conducted on ground powders to remove possible surface effects in pelleted samples.

\subsection{X-ray fluorescence microscopy}

X-ray fluorescence microscopy studies were conducted using the x-ray nanoprobe at beamline ID21 of the European Synchrotron Radiation Facility ~\cite{ID21-ref}. Measurements were conducted in total fluorescence yield (TFY) mode, with a beam size of 0.3 $\mu$m in the vertical direction, and 0.8 $\mu$m in the horizontal direction. Data was collected for the Cr, Mn, Fe, and Co K$_\alpha$ fluorescence lines. Ni fluorescence data was unattainable due to design considerations, as the beamline's optics utilize Ni-coated mirrors.

\subsection{Scanning electron microscopy and energy dispersive x-ray spectroscopy}

Electron micrographs were obtained using a Zeiss Crossbeam350 FIB-SEM microscope. Images were acquired using both secondary electron imaging (SEI) and backscattered electron (BSE) data channels. Energy dispersive x-ray spectroscopy (EDX) measurements were performed on a Philips XL30 scanning electron microscope equipped with a Bruker Quantax 200 energy-dispersion x-ray microanalysis system and an XFlash 6010 SDD detector.

\subsection{X-ray absorption and magnetic circular dichroism}

X-ray magnetic circular dichroism (XMCD) is a subset of x-ray absorption (XAS), which achieves magnetic sensitivity by irradiating a sample with left and right-handed circularly polarized light, measuring the corresponding absorption spectra, and subtracting them from one another \cite{Thole-Sawatzky-XMCD-prediction, VanderLaan-Sawatzky-XMCD-proof}. The existence of circular dichroism requires time-reversal symmetry breaking, as observed in a magnetic material. Furthermore, being a core electron spectroscopy, XMCD inherits XAS's elemental specificity and crystal field and oxidation state sensitivities. 

XAS and XMCD experiments were carried out at the elastic endstation of the REIXS beamline of the Canadian Light Source \cite{REIXS-beamline}. The experiment was carried out at normal incidence, with the sample temperature being kept at 20 K throughout the data collection. The spectra were measured using an energy resolving silicon drift detector, and by measuring the drain current when possible. Inverse partial fluorescence yield spectra were obtained by monitoring the oxygen fluorescence signal while scanning across the transition metal L$_{2,3}$ edges. IPFY measurements have the advantage of being bulk-sensitive while avoiding the self-absorption effects that plague conventional fluorescence detection methods \cite{Achkar-Hawthorn-IPFY}. The samples were mounted on silver paint to improve thermal contact.

The XAS and XMCD measurements were then contrasted with ligand field multiplet theory calculations performed using the Quanty code \cite{lanczos1950iteration,Quanty}. Detailed descriptions and lists of parameters for the simulations are provided in the Supporting Information. 

\subsection{Extended x-ray absorption fine structure}
Extended x-ray absorption fine structure (EXAFS) measurements were carried out at the hard x-ray microanalysis (HXMA) beamline of the Canadian Light Source. Samples were ground until a homogenous powder was obtained and then spread thinly and homogenously on Kapton tape. The corresponding elemental metal foils were used for energy calibration at each edge.  Measurements were then carried out in transmission and fluorescence yield modes at ambient pressure and temperature. The data were then post-processed and exported using the Athena package of the Demeter Suite ~\cite{Ravel-DemeterSuite}.

The predominant physical process in x-ray absorption spectroscopy (XAS) is for a photoelectron to be ejected, leaving a core-hole. Due to its wave-like nature, this photoelectron is free to backscatter off the neighboring atoms, and the resulting wave can in turn interact with itself. This self-interference process gives rise to extended x-ray absorption fine structure (EXAFS). EXAFS manifests as oscillatory modulations superimposed on the constant background of normalized XAFS spectra. These modulations are then a signature of the first couple of coordination shells around the absorber.

\subsection{Magnetic susceptibility}

Temperature and field-dependent magnetic susceptibility measurements were performed using a Quantum Design MPMS3 SQUID magnetometer. All samples were mounted in gel capsules. Susceptibility measurements were then performed under zero field-cooled and field-cooled conditions between 2 and 400 K, under an applied field H = 0.01 T. To reach the paramagnetic state, measurements were then extended to 600 K using the Oven option. For the magnetization isotherms, data was collected at 2, 100, and 200 K for fields between -3 and 3 T. In all cases, samples were heated about their magnetic ordering transition temperatures and cooled under zero-field conditions between magnetization isotherms were measured. 

\section*{Supporting Information}
The Supporting Information contains (i) Additional figures for x-ray diffraction, scanning electron microscopy, microscopically resolved x-ray fluorescence, extended x-ray absorption fine structure, and x-ray absorption; (ii) Parameters used in the models for the multiplet calculations, and fits to the XAS data; and (iii) Tables containing quantitative values for structural and magnetic parameters of the samples studied here.

\begin{acknowledgments}

The authors express their appreciation to Uwe Engelhardt and Frank Falkenberg at the Max Planck Institute for Solid State Research, who provided experimental support for the high pressure synthesis. From the University of British Columbia, the authors acknowledge Andrea Ortiz for experimental support with collecting SEM data, Graham Johnstone, Christine Trinh, and Morgan Brand for solid state synthesis, and Benjamin Herring for experimental support on TGA measurements. The authors thank Andrea Damascelli, Joerg Rottler, and George Sawatzky for insightful conversations on high entropy oxides. This work was funded in part by a QuantEmX grant from ICAM and the Gordon and Betty Moore Foundation through Grant GBMF9616 to Solveig Stubmo Aamlid. This work was supported by the Natural Sciences and Engineering Research Council of Canada (NSERC), the CIFAR Azrieli Global Scholars program, and the Sloan Research Fellowships program. This research was undertaken thanks in part to funding from the Max Planck-UBC-UTokyo Center for Quantum Materials and the Canada First Research Excellence Fund, Quantum Materials and Future Technologies Program. Part of the research described in this paper was performed at the Canadian Light Source, a national research facility of the University of Saskatchewan, which is supported by the Canada Foundation for Innovation (CFI), the Natural Sciences and Engineering Research Council (NSERC), the National Research Council (NRC), the Canadian Institutes of Health Research (CIHR), the Government of Saskatchewan, and the University of Saskatchewan. 
\end{acknowledgments}

\bibliography{bibliography}

%\begin{figure*}
 % \centering
  %\includegraphics[width=3.25in]{JACS_TOC.eps}
  %\caption{For Table of Contents Only}
  %\label{fgr:TOC}
%\end{figure*}

\end{document}